\documentstyle[12pt]{article}

\newcommand{\EQ}{\begin{equation}}
\newcommand{\EN}{\end{equation}}
\newcommand{\bea}{\begin{eqnarray}}
\newcommand{\ena}{\end{eqnarray}}
\newcommand{\vs}[1]{\vspace{#1 mm}}

\newcommand{\uda}{\nearrow \kern-1em \searrow}

\begin{document}

\topmargin 0pt
\oddsidemargin 5mm

\renewcommand{\Im}{{\rm Im}\,}
\newcommand{\NP}[1]{Nucl.\ Phys.\ {\bf #1}}
\newcommand{\PL}[1]{Phys.\ Lett.\ {\bf #1}}
\newcommand{\CMP}[1]{Comm.\ Math.\ Phys.\ {\bf #1}}
\newcommand{\PR}[1]{Phys.\ Rev.\ {\bf #1}}
\newcommand{\PRL}[1]{Phys.\ Rev.\ Lett.\ {\bf #1}}
\newcommand{\PTP}[1]{Prog.\ Theor.\ Phys.\ {\bf #1}}
\newcommand{\PTPS}[1]{Prog.\ Theor.\ Phys.\ Suppl.\ {\bf #1}}
\newcommand{\MPL}[1]{Mod.\ Phys.\ Lett.\ {\bf #1}}
\newcommand{\IJMP}[1]{Int.\ Jour.\ Mod.\ Phys.\ {\bf #1}}
\begin{titlepage}
\setcounter{page}{0}
\begin{flushright}
OU-HET 264\\
hep-ph/9705263\\
\end{flushright}
\vs{15}
\begin{center}
{\Large The rephasing freedom and the NNI form of quark mass matrices}
\vs{15}

{\large Eiichi Takasugi}\\
\vs{8}
{\em Department of Physics, Osaka University \\ 
 1-16 Toyonaka, Osaka 560, Japan} \\
\end{center}
\vs{10}
\centerline{{\bf{Abstract}}} 
For three generations of quarks, we show that  
 one of  quark mass matrices can 
be transformed   into either a Fritzsch form or  
a Branco-Silva-Marcos form, 
while the other is kept in the NNI basis. 
In these bases, quark mass matrices are determined 
unambiguously  
once quark masses and the CKM mixing are given. 
\end{titlepage}
\newpage
\renewcommand{\thefootnote}{\arabic{footnote}}
\setcounter{footnote}{0}

\section{Introduction}

Since Fritzsch[1] introduced an so called Fritzsch ansatz 
for quark mass matrices, the understanding of the meaning 
of this ansatz has been an important issue in the 
research on finding the origin of the quark mixing. 
Branco, Lavoura and Mota[2] have shown that quark mass 
matrices can be transformed into the NNI 
(nearest neighbor interaction) form without loss of 
generality. They argued that  the Fritzsch ansatz is 
realized if the hermiticity is required on masses in 
the NNI basis. However, it is known that the Fritzsch 
ansatz for both u- and down-quark mass matrices does not explain 
the experimental data of quark mixing.  
Branco and Silva-Marcos[3] proposed another 
ansatz in the NNI form to which we refer as a BS 
form. The BS form has a complimentary feature to 
the Fritzsch form for the quark mixing and 
seems to be another important 
ansatz. Recently, Ito[4] showed that the experimental 
data of the quark mixing are explained well if the 
Fritzsch ansatz is used for the up-quark mass matrix 
$M_u$ and the BS ansatz for the down-quark mass matrix. 
However, there is no reasonable explanation 
of the Branco ansatz up to now.   

Recently, Koide[5] reported an interesting observation that 
for three generations, 
the 3-2 element of the up-quark mass matrix $M_{u32}$ can 
be made zero by using the rephasing freedom of quarks 
in the NNI basis.  In this paper, 
inspired by Koide's work[5], we investigate 
further the role of the rephasing freedom of quarks and 
its relation to the transformation explicitly. 
By using the rephasing freedom, we constructed explicitly  
the transformation of  $M_u$ into   
the Fritzsch form while $M_d$ is kept in  the NNI form 
(hereafter we refer to it as 
the Fritzsch-type basis) and also 
the transformation of $M_d$ into the BS form while $M_u$ is kept 
in the NNI basis (hereafter we refer 
to it as the BS-type basis). 
For three generations, the rephasing freedom contains 
two phase parameters which are adjusted to make these 
transformations. Phase parameters are determined by a 
complex equation.  We analyzed this equation  
and found out that there exist solutions of phase parameters 
for reasonable mass matrices such that the experimental values of 
the quark 
mixing are reproduced.  Since we deal with  
reasonable mass matrices in practice, these transformations exist 
for all practically meaningful matrices. 
As a result,  for three generations  
the  Fritzsch ansatz for $M_u$ or the BS ansatz for $M_d$ 
in the NNI basis is 
not the ansatz anymore, but these can be realized by the transformation 
that leaves the CKM matrix invariant. 

Let us define the problem clearly. Firstly, 
we define mass terms as $\bar u_LM_uu_R+\bar d_LM_dd_R$ 
and  the unitary transformation  for left-handed quarks and 
the right-handed quarks which leaved the CKM quark mixing matrix 
invariant:
\bea
\tilde U^{\dagger}M_u\tilde V_u&=&\tilde M_u\;,\nonumber\\
\tilde U^{\dagger}M_d\tilde V_d&=&\tilde M_d\;.
\ena
Here $\tilde U$ is the transformation matrix of 
left-handed quarks, 
$\tilde V_q (q=u,d)$ are transformation matrices of 
right-handed quarks. 
Branco et.al.[2] showed that by this 
transformation, $\tilde M_q$ can be made into the NNI basis,  
\bea
\tilde M_q=\left (\matrix{0 &a_q&0\cr
                          b_q&0&c_q\cr
                          0&d_q&e_q\cr}\right ).
\ena
Our question is to seek the possibility to transform $M_u$ or $M_d$ into 
some special forms by keeping the other mass matrix in the NNI form. 
Koide[5] pointed out that there is the rephasing freedom of quarks 
which we use for the above transformations.  
The rephasing freedom can be regarded in a different 
way. The unitary matrices for the transformation in 
Eq.(1) are $\tilde U$, $\tilde V_u$ and $\tilde V_d$. 
Each unitary matrix has six freedoms aside from unimportant 
phase freedoms for every column. Thus  there are 
eighteen freedoms in the transformation. The requirement 
for the NNI form for mass matrices amounts to sixteen 
real constraints. Thus, there remain two freedoms which 
correspond to the rephasing freedom. Thus, we really 
treat the transformation in Eq.(1). 
This freedom is present only for three generations 
of quarks. 

In Sec.2, we see how $M_u$ can be transformed to a 
Fritzsch form or how $M_d$ can be transformed to a 
BS form. We obtain a complex equation 
containing two phases for each case. In Sec.3, we show that 
these equations have solutions for two phases, when 
mass matrices are ones which reproduce the CKM quark 
mixing. Summary is given in Sec.4.  

\section{ Transformation  to the  Fritzsch-type or the BS-type basis}

Let us start by diagonalizing quark mass matrices as 
\bea
 U_u^{\dagger}M_uV_u =D_u\;,
 U_d^{\dagger}M_dV_d =D_d\;,
\ena
where $D_q$ is a diagonal matrix. The CKM matrix $K$ is expressed by 
\bea
K=P_{u}^{\dagger}U_u^{\dagger}U_dP_{d}\;,
\ena
where $P_{q} $  are diagonal phase matrices originated from 
the rephasing freedoms of quarks.  The important observation is 
that $P_{q}$ can not be fixed yet because  they change if 
the phases of column vectors in  unitary matrices $U_q$ and 
$V_q$ are changed. Therefore, we consider $P_q$ are arbitrary 
phase matrices. This phase freedom is called the rephasing freedom 
by Koide[5]. 

Now we rewrite transformations in similar forms in Eq.(1) as 
\bea
U_u^{\dagger}M_uV_u=D_u\;,\quad 
U_u^{\dagger}M_dV_d=P_uKP_d^{\dagger}D_d\;,
\ena
or
\bea
U_d^{\dagger}M_uV_u=P_dK^{\dagger}P_u^{\dagger}D_u\;,\quad 
U_d^{\dagger}M_dV_d=D_d\;.
\ena
For both forms, the transformation matrix for the left-handed 
quarks is the same for the up- and down-quarks so that the 
CKM matrix is left invariant.

\vskip 3mm
\noindent
(a) The  Fritzsch-type basis

Here, we shall transform the diagonal basis of mass matrices to 
the one  where $M_u$ takes the Fritzsch form as 

\bea
(U_uO_u^T)^{\dagger}M_u(V_uO_u^T)=O_uD_uO_u^T=\tilde M_{uF}\;,\nonumber\\
(U_uO_u^T)^{\dagger}M_d(V_dV_d')=O_uP_uKP_d^{\dagger}D_d V_d'=\tilde M_d\;.
\ena
We used Eq.(5) to derive the above equation. 
In order for $\tilde M_{uF}$ to be a Fritzsch  matrix 
\bea
\tilde M_{uF}=\left (\matrix{0 &a_u&0\cr
                          a_u&0&c_u\cr
                          0&c_u&e_u\cr}\right )\;,                        
\ena
we chose $O_u$ to be  the orthogonal matrix which  
diagonalizes the $\tilde M_{uF}$. Here 
$a_u$, $c_u$ and $e_u$ are all real parameters and are 
expressed by the up-quark masses, $m_u$, $m_c$ and $m_t$. 
Similarly,   the 
orthogonal matrix $O_u$ is expressed uniquely by these 
quark masses.  $M_d$ is transformed 
by introducing the unitary matrix $V_d'$ for the right-handed 
down-quarks  which is utilized 
to transform  $\tilde M_d$  into a NNI form. 
 
If  $\tilde M_d$ can be transformed  into the NNI form, then   
 the unitary matrices to transform into the Fritzsch-type basis are 
 readily obtained.  Explicitly, 
by comparing Eq.(7) with Eq.(1), we find    $\tilde U=U_uO_u^T$, 
$\tilde V_u=V_uO_u^T$ and $\tilde V_d=V_dV_d'$.  The transformation 
for $M_d$ contains a unitary matrix $V_d'$ and 
the phase matrix $P_u$ which  will play an important role. 

Now the question is 
how  $\tilde M_d$ can  be made into  the NNI form, 
\bea
 \tilde M_d=\left (\matrix{0 &a_d e^{i\alpha_1}&0\cr
                          b_d&0&c_d e^{i\alpha_2}\cr
                          0&d_d&e_d\cr}\right)\;,
\ena
where $a_d$, $b_d$, $v_d$, $d_d$ and $e_d$ are 
real nonnegative numbers. 
As proved  by Branco et al.[2], if  the condition 
\bea 
H_{d12}=0\;
\ena
is satisfied, $\tilde M_d$ can be transformed into 
the NNI form by taking an appropriate $V_d'$. Here, 
 $H_d$ is an hermitian matrix defined by 
\bea
H_d\equiv \tilde M_d\tilde M_d^{\dagger}=O_uP_u(KD_d^2K^{\dagger})
P_u^{\dagger} O_u^T \;.
\ena

The condition $H_{d12}=0$ contains only phases in $P_u$ 
as adjustable  parameters  because  $(KD_d^2K^{\dagger})$ is written 
only by the observable, masses of quarks and CKM matrix 
elements. The condition is expressed explicitly by
\bea
\sum_{j,k=1,2,3}e^{i(\theta_j-\theta_k)}(O_u)_{1j}(O_u)_{2k}
(KD_d^2K^{\dagger})_{jk}=0\;,
\ena
where we used $P_u={\rm diag}(\exp(i\theta_1),\exp(i\theta_2),
\exp(i\theta_3))$. The above complex equation contains 
two independent phases $\theta_1-\theta_2$ and  
$\theta_2-\theta_3$ so that in general this equation 
is satisfied by taking appropriate values of these phases.  
In the next section, we see in detail that this equation 
is satisfied by taking appropriate values of $\theta_1-\theta_2$ and  
$\theta_2-\theta_3$.  

The NNI form $\tilde M_d$ is given by $H_d$ as  
\bea
a_d&=&\sqrt{H_{d11}}\;,\quad \; b_d=\sqrt{\frac{\det H_d}{
H_{d11} H_{d33}- |H_{d13}|^2}}\;,\nonumber\\
c_d&=&\frac{H_{d23}\sqrt{ H_{d11}}}{\sqrt{ H_{d11}
 H_{d33}-|H_{d13}|^2}}\;,\quad \;
d_d=\frac{ |H_{d13}|}{ \sqrt{H_{d11}}}\;,\nonumber\\
e_d&=&\frac{\sqrt{ H_{d11} H_{d33}- |H_{d13}|^2}}
   {\sqrt{H_{d11}}}\;,\nonumber\\
\alpha_1&=&\arg H_{d13}\;, \quad \alpha_2=\arg H_{d23}\;,
\ena
where $H_{d11}$, $H_{d22}$, $H_{d33}$ and $\det H_{d}$ are 
real nonnegative numbers. 
Now the up-quark mass matrix is fixed only by up-quark masses 
and the down-quark mass matrix is given by quark masses and 
the CKM elements. As a result, quark mass matrices are expressed 
explicitly by the observable. Recently, Harayama and Okamura[6] 
investigated the inverse problem to express the quark mass 
matrices in terms of the observable. They used the NNI form 
of quark mass matrices which contains twelve parameters 
and thus the quark mass matrices contain two free parameters. The 
fixing of these parameters was a problem. Koide[5] 
succeeded to remove these freedoms by transforming into a special 
basis where $M_{u32}=0$ in the NNI form, but this form was 
not symmetric. Our bases are symmetric ones and 
mass matrices are expressed unambiguously by the  observable 
so that this form will be useful  for fixing the quark 
mass matrices in practice. 

\vskip 3mm
\noindent 
(b) The BS-type basis  
 
The procedure to transform $M_d$ into the BS form is 
essentially the same as the Fritzsch case. We make 
the following transformation,  
\bea
 (U_dO_d^T)^{\dagger}M_d(V_dO_d'^T)=O_dD_uO_d'^T=\tilde M_{dB}\;,\nonumber\\
 (U_dO_d^T)^{\dagger}M_u(V_uV_u')=O_dP_dK^{\dagger}P_u^{\dagger}
 D_u V_u'=\tilde M_u\;,
\ena
where $ M_{dB}$ is a BS form defined by 
\bea
\tilde M_{dB}=\left (\matrix{0 &a_d&0\cr
                          a_d&0&c_d\cr
                          0&e_d&e_d\cr}\right )\;   
\ena
and $O_d$ is the orthogonal matrix which diagonalizes 
$\tilde M_{dB}$ as $O_d^T\tilde M_{dB}O_d'=D_d$. 
Here, $a_d$, $c_d$ and $e_d$ are all real parameters and $\tilde M_u$ is an 
complex matrix.  

If $\tilde M_u$ is made into the NNI form, the transformation 
in Eq.(1) is given by taking  $\tilde U=U_dO_d^T$, 
$\tilde V_d=V_dO_d'^T$ and $\tilde V_u=V_uV_u'$.  
 $\tilde M_u$ becomes the NNI form  by using the 
freedoms of $P_d$  and $V_u'$ once the condition 
\bea
( H_u)_{12}=0\;
\ena
is satisfied, where
\bea
H_u=\tilde M_u\tilde M_u^{\dagger}=O_dP_d(K^{\dagger}D_u^2K)
 P_d^{\dagger} O_d^T \;.
\ena
The requirement of 
$(H_u)_{12}=0$ is 
\bea
\sum_{j,k=1,2,3}e^{i(\phi_j-\phi_k)}(O_d)_{1j}(O_d)_{2k}
(K^{\dagger}D_u^2K)_{jk}=0\;,
\ena
where we used $P_d={\rm diag}(\exp(i\phi_1),\exp(i\phi_2),
\exp(i\phi_3))$. We show in the next section that 
by choosing appropriate values of  two independent phases 
$\phi_1-\phi_2$, 
$\phi_2-\phi_3$, the above equation is satisfied. 
When we write $\tilde M_u$ as 
\bea
 \tilde M_u=\left (\matrix{0 &a_u e^{i\beta_1}&0\cr
                          b_u&0&c_u e^{i\beta_2}\cr
                          0&d_u&e_u\cr}\right)\;,
\ena
then each element and phases are  expressed in terms of $H_u$ 
as in Eq.(13). 
In this basis, quark mass matirices $\tilde M_{dS}$ and $\tilde M_u$ 
can be expressed uniquely by quark masses and CKM mixing.

\section{Examination of  constraint equations}

In order to examine  constraint equations in Eqs.(12) and (18), we 
need to parameterize the CKM matrix for which 
 we use the following form, 
\bea
K\simeq \left(\matrix{1-\frac 12 \lambda^2& \lambda&\sigma e^{-i\delta}\cr
                     -\lambda& 1-\frac 12 \lambda^2& \rho\cr
                     \sigma' e^{i\delta'}&-\rho&1\cr
                      }  \right)\;,
\ena 
where 
$\sigma'=|K_{td}|$ and phase $\delta'$ are given by  
\bea
\sigma' &=&\sqrt{(\lambda\rho)^2+\sigma^2-2\lambda\rho\sigma\cos \delta}\;,
\nonumber\\
\sigma'\cos \delta'&=&\lambda\rho-\sigma\cos\delta\;,\quad
\sigma'\sin\delta'=-\sigma\sin\delta\;.
\ena  
Here, the orders of magnitudes of parameters in CKM matrix are 
$\rho=|K_{cb}|\sim O(\lambda^2)$, $\sigma=|K_{ub}|\sim (\lambda^4)$ and
$\sigma'=|K_{td}|\sim O(\lambda^3)$. 
We also define the ratios of quark masses as 
\bea
r_d&\equiv& m_d/m_s\sim O(\lambda^2)\;, 
\quad r_s\equiv m_s/m_b\sim O(\lambda^{5/2})\;,
\nonumber\\ 
r_u&\equiv& m_u/m_c\sim O(\lambda^4)\;, \quad 
r_c\equiv m_c/m_t\sim O(\lambda^4)\;. 
\ena
By using these parameters,  conditions are examined. 

\vskip 2mm
\noindent
(a) The case of the  Fritzsch-type basis

Firstly, we evaluate hermitian matrix $KD_d^2K^{\dagger}$ by 
using quark masses and CKM parameters in the leading order 
of $\lambda$. The result is as follows: 
\bea 
(KD_d^2K^{\dagger})_{11}&\simeq& (\sigma^2 +(r_s\lambda)^2)m_b^2 \;,\nonumber\\
(KD_d^2K^{\dagger})_{22}&\simeq& (\rho^2 +r_s^2)m_b^2 \;,\nonumber\\
(KD_d^2K^{\dagger})_{33}&\simeq& m_b^2\;,\nonumber\\
(KD_d^2K^{\dagger})_{12}&\simeq& (\rho\sigma e^{-i\delta}+r_s^2\lambda)m_b^2 
\;,\nonumber\\
(KD_d^2K^{\dagger})_{13}&\simeq& \sigma e^{-i\delta} m_b^2 \;,\nonumber\\
(KD_d^2K^{\dagger})_{23}&\simeq& \rho m_b^2 \; .
\ena
The orthogonal matrix which diagonalizes the Fritzsch mass matrix for 
up-quarks is given by 
\bea
O_u\simeq\left(\matrix{1&-\sqrt{r_u}&r_c\sqrt{r_ur_c}\cr
                  \sqrt{r_u}&1&\sqrt{r_c}\cr
                  -\sqrt{r_ur_c}&-\sqrt{r_c}&1\cr}  \right)
\ena
By keeping leading order terms, the condition in Eq.(12) is 
expressed as  
\bea
Ae^{i\theta_{12}}-Be^{i\theta_{23}}+Ce^{i(\theta_{12}+
\theta_{23}-\delta)}=D\;,
\ena
where $\theta_{12}=\theta_1-\theta_2$, $\theta_{23}=\theta_2-\theta_3$
\bea
A\equiv |A|e^{-i\kappa}=\rho\sigma e^{-i\delta}+r_s^2\lambda \;, 
B=\sqrt{r_ur_c}\rho \;,
C=\sqrt{r_c}\sigma \;,  
D=\sqrt{r_u}(\rho^2+r_s^2)\;.
\ena
It is worthwhile to note that $B$, $C$, $D$ are real positive numbers, 
while $A$ is a complex number. 
By taking the absolute values of both   sides of 
$e^{i(\theta_{12}-\kappa)}(|A|+Ce^{i(\theta_{23}-\delta+\kappa)})
=Be^{i\theta_{23}}+D$, we find
\bea
2[BD\cos \theta_{23}-|A|C\cos(\theta_{23}-\delta+\kappa)]=
|A|^2+C^2-B^2-D^2\;.
\ena
This equation has a solution if the following inequality is satisfied, 
\bea
\left| \frac{|A|^2+C^2-B^2-D^2}{2\sqrt{B^2D^2+|A|^2C^2-2|A|BCD
\cos(\delta-\kappa')}} \right|\le 1\;.
\ena
This inequality constrains the CP violation angle $\delta$ and 
the absolute values of CKM matrix elements 
$\lambda$, $\rho$ and $\sigma$ are given. The condition is simply 
written as  
\bea
\alpha\cos^2\delta+2\beta\cos\delta+\gamma\le 0\;,
\ena
where
\bea
\alpha&=&4(r_s^2\lambda\rho\sigma)^2\;,\nonumber\\
\beta&=&2r_s^2\lambda \{\rho\sigma [(r_s^2\lambda)^2+(\rho\sigma)^2
     +C^2-B^2-D^2] +2C(BD-\rho\sigma C)\}\;,\nonumber\\
\gamma&=&[(r_s^2\lambda)^2+(\rho\sigma)^2+C^2-B^2-D^2]^2
   -4[(BD-\rho\sigma C)^2
  +(r_s^2\lambda C)^2]\;.
\ena

We consider the above equation as the constraint equation for 
the CP violation angle $\delta$ by giving the experimental 
values of $\lambda$, $\rho$ and $\sigma$. There are some 
uncertainties in the experimental values of them. Here we take 
the central values to see in what region of $\delta$,  
there exist solutions. 
We use the CKM parameters[7], $\lambda\equiv |K_{us}|=0.2205$, 
$\rho\equiv |K_{cb}|=0.041$ and $\sigma/\rho\equiv |K_{ub}|/|K_{cb}|
=0.08$ and the running  quark masses defined at $\mu=M_Z$, 
$m_u=0.00222$GeV, $m_d=0.00442$GeV, $m_s=0.0847$GeV, $m_c=0.661$GeV, 
$m_b=2.996$GeV and $m_t=180$GeV. Then, the constraint gives  
$\cos \delta \le 0.66$. 
This means that for reasonable varieties  of 
quark mass matrices which reproduce the CKM matrix, 
the up-quark mass matrix can be transformed to the Fritzsch 
form, while the down-quark mass matrix is kept in the NNI form. 
It is necessary to examine further the region of parameters 
of CKM matrix which allows this transformation. 

\vskip 2mm
\noindent
(b) The case of  the BS-type basis 

Analysis can be done similarly to the Fritzsch case. We find 
\bea 
(K^{\dagger}D_u^2K)_{11}&\simeq& \sigma'^2 m_t^2\;,\nonumber\\
(K^{\dagger}D_u^2K)_{22}&\simeq& \rho^2 m_t^2 \;,\nonumber\\
(K^{\dagger}D_u^2K)_{33}&\simeq& m_t^2\;,\nonumber\\
(K^{\dagger}D_u^2K)_{12}&\simeq&-\rho\sigma'e^{-i\delta'}m_t^2 \;,\nonumber\\
(K^{\dagger}D_u^2K)_{13}&\simeq& \sigma' e^{-i\delta'} m_t^2 \;,\nonumber\\
(K^{\dagger}D_u^2K)_{23}&\simeq& -\rho m_t^2\;.
\ena
Then,  by using 
\bea
O_d\simeq\left(\matrix{1&-2^{-1/4}\sqrt{r_d}& 
         2^{-1/4}r_s\sqrt{r_d}\cr
          2^{-1/4}\sqrt{r_d}&1&r_s\cr
           -2^{3/4}r_s\sqrt{r_d}&-r_s&1\cr}
 \right)
\ena
by keeping leading terms, we 
find that the condition is reduced to 
\bea
A'e^{i(\phi_{12}-\delta')}-iB'\sin\phi_{23}-C'e^{i(\phi_{12}-\delta'+
\phi_{23})}=-D'\;, 
\ena
where $\phi_{12}=\phi_1-\phi_2$, $\phi_{23}=\phi_2-\phi_3$ and 
\bea
A'=\rho \sigma' \;, 
\quad B'=2^{3/4}r_s\sqrt{r_d}\rho \;,\quad C'=r_s\sigma'\;,\quad  
D'=2^{-1/4}\sqrt{r_d}(\rho^2-r_s^2) \;.
\ena
By taking the absolute values of both sides of  the equation 
$e^{i(\phi_{12}-\delta')}(A'-C'e^{i(\phi_{23})})=iB'\sin\phi_{23}-D'$, 
we obtain the equation for $\phi_{23}$ as
\bea
B'^2\cos^2\phi_{23} -2A'C'\cos\phi_{23}+A'^2+C'^2-B'^2-D'^2=0\;.
\ena

By using the values of quark masses and CKM elements given  
before and 
$\sigma'\equiv |K_{td}|=0.009$[7], we obtain $\cos\phi_{23} =-0.12$. 
The phase $\phi_{12}$ is determined once $\phi_{23}$ 
is given. Although we did not 
take into account the experimental uncertainties in CKM elements, 
 the above analysis  suggests  that for reasonable quark mass matrices, 
 the condition will be satisfied. 
Therefore,  
it is possible that 
the down-quark mass matrix can be transformed into the BS 
form, while the up-quark mass matrix is kept in the NNI form 
in practice.

\section{Summary}
For three generation of quarks, there still remain two freedoms of 
the transformation of quark mass matrices which leave the CKM matrix 
invariant in the NNI basis. 
By utilizing these freedoms, we showed that either the Fritzsch-type 
or the BS-type basis is possible. Here the Fritzsch-type 
means that $M_u$ takes the Fritzsch form, while $M_d$ does 
 the NNI form. On the other hand.  The BS-type means that $M_d$ takes the 
BS form, while $M_u$ does  the NNI form. 
We explicitly constructed these transformations and saw that the rephasing 
freedom 
of quarks plays an important role. 
Our work is an extension of Koide's basis[5] where $M_{u32}=0$ in the 
NNI basis. 

The two bases, the Fritzsch-type and 
the BS-type  as well as Koide's basis contain only ten physical 
parameters in quark mass matrices so that quark mass matrices are 
expressed only by the observable, quark masses and CKM parameters. 
Both of our bases are symmetric type parameterizations contrary to Koide's one 
which is asymmetric type. 

It may be worthwhile to see 
the transformation between the Fritzsch-type and the BS-type bases 
which is explicitly given by 
\bea
\tilde M_u&=&(O_uP_uKP_d^{\dagger}O_d^T)^T\tilde M_{uF}(O_uV_u')\;,
\nonumber\\
\tilde M_{dB}&=&(O_uO_d^T)^T\tilde M_d(O_uO_d'^T)\;.
\ena
The transformations between Koide's basis and our ones are 
similarly obtained, but we do not list here. 
 
It is interesting to observe that  the Fritzsch-BS ansatz where $M_u$ takes 
the Fritzsch form and $M_d$ does the BS form can be obtained 
from the BS basis by requiring the hermiticity for $M_u$. 
As discussed by Ito[4], the Fritzsch-BS ansatz may well be an good 
ansatz to explain the CKM matrix if the CP phase is in the first 
quadrant in the $\rho-\eta$ plane. 

There are several works left. One is the examination on the region 
of CKM parameters which allows these transformations. The other is 
to express quark mass matrices  by observable,  quark masses and CKM 
parameters. These works will be reported in the forthcoming 
paper[8].

\end{document}